\documentclass{article}
\usepackage[utf8]{inputenc}
\usepackage{physics,amsmath,amsfonts,stmaryrd,tikz}
\usepackage{url}

\newcommand{\ita}{\textit}
\newcommand{\mcal}{\mathcal}

\newcommand{\mbf}{\mathbf}

\newcommand{\lra}{\leftrightarrow}

\title{The Non-Individuals Interpretation of Quantum Mechanics \\ }
\author{Décio Krause \and Jonas R. B. Arenhart \and Otávio Bueno}

\DeclareUnicodeCharacter{2212}{-}

\begin{document}
\maketitle

\section{Introduction}

In this chapter, we present a `non-individuals' interpretation of quantum mechanics. This is not among the well-known interpretations of quantum theory, such as Bohmian mechanics, the many-worlds interpretation, and GRW. Talk of non-individuals in quantum mechanics concerns primarily the metaphysics of the theory (French and Krause \cite{fre06} and French \cite{fre19}). However, as we argue, there is a clear sense in which non-individuality, as a metaphysical concept, is associated with interpreting quantum mechanics.

Our arguments establish salient connections between interpreting quantum theory and the metaphysics of quantum individuality. We specify the meaning of non-individuality and relate it to current efforts regarding the interpretation of quantum theory. We begin by recalling that an interpretation of quantum mechanics, as currently understood, is related to solutions to the measurement problem. Such solutions, we claim, are directly connected to attempts at providing a picture of what the world looks like according to quantum mechanics. In particular, we shall advance the claim that an interpretation, in the relevant sense, brings with it an ontology: an account of what the furniture of the world is like, according to quantum mechanics. This step takes us from interpretation to ontology. The shift from ontology to non-individuality is just another, albeit controversial, further step. Philosophical interest in ontology is typically accompanied by interest in articulating a deeper understanding of the nature of the relevant entities, and a proposal for a metaphysics of non-individuals finds its way at this point. If quantum mechanics is understood as dealing with objects of a given kind, whether particles, fields or something else, it may be asked: what metaphysically are these objects? This, in turn, leads to questions regarding whether they are individuals or not, and if they are, which principle of individuality determines that that is the case? The non-individuals interpretation of quantum mechanics takes the relevant entities as being devoid of individuality, adding a further metaphysical interpretative layer over the theory's bare entities. This is known as \textit{the Received View} of quantum non-individuality (see French and Krause \cite[Chapter 3]{fre06} for a historical overview).

This only tells us that a metaphysical view is attached to the ontology of an interpretation, but it does not specify what non-individuals are, or how the relevant concepts should be understood. Once a metaphysical interpretation is connected with an ontology, the task of making the metaphysics explicit still remains. Considered at this level of generality, the Received View is only a scheme, rather than a specific thesis. It acquires metaphysical content only once it is specified what an individual is and how something may fail to be one---that is, once an articulation of the Received View in metaphysical terms is provided. We examine two ways of articulating the view: a metaphysically inflationary form, and a metaphysically deflationary way.

The Received View, as a thesis about the nature of quantum entities, has its historical roots in the work done by those who first formulated quantum mechanics. It originates from attempts to understand the nature of the entities described by quantum theory. In particular, quantum particles' lack of well-defined trajectories and the presence of permutation invariance contrasted strikingly with the behavior of classical particles (French and Krause \cite[Chapters 2 and 3]{fre06}). In this way, the Received View, even if considered as adding a metaphysics over the entities posited by an interpretation of quantum mechanics, has a historical pedigree.

This chapter is structured as follows. We begin, in the next section, by connecting an interpretation of quantum mechanics---understood as aiming to address the measurement problem---and an ontology. The plan is to make room to interpret further some entities in metaphysical terms. In Section \ref{sec3} we discuss how a metaphysical profile can be attached to such entities, and how a philosophical motivation is typically offered for such move. We then briefly examine, in Section \ref{sec4}, the main concepts involved in articulating a metaphysics of non-individuals. This is followed, in Section \ref{sec5}, by the presentation of what is perhaps the most well-known metaphysics of non-individuals: one in which the concept of identity has no place. In Section \ref{sec6} we analyze a less revisionary alternative to the standard account. Finally, in Section \ref{sec7}, a conclusion is offered.

\section{Interpretation and ontology}\label{sec2}

The idea that a physical theory needs an interpretation is rather recent in the history of physics, appearing after the development of modern physics itself (see also the discussion in Mittelstaed \cite{mit11}). Independently of interpretational issues in quantum mechanics, the very notions of theory and interpretation remain subjects of philosophical controversy, but we will not pursue this here (see Krause and Arenhart \cite{kra17}). Following Esfeld \cite{esf19}, we take the fact that quantum theory needs an interpretation to be part of the very aim of that physical theory. As Esfeld puts it:
\begin{quote}
[\dots] a physical theory has to (i) spell out an ontology of what there is in nature according to the theory, (ii) provide a dynamics for the elements of the ontology and (iii) deduce measurement outcome statistics from the ontology and dynamics by treating measurement interactions within the ontology and dynamics; in order to do so, the ontology and dynamics have to be linked with an appropriate probability measure. Thus, the question is: What is the law that describes the individual processes that occur in nature (dynamics) and what are the entities that make up these individual processes (ontology)? (Esfeld \cite[p. 222]{esf19})
\end{quote}

As is well known, to find an appropriate account of the dynamics has been a key task to ensure the proper formulation and understanding of quantum mechanics (dynamical laws are crucial to characterize the theory). This is also a source of difficulty in theory individuation: formulations of quantum theory that differ in their dynamics may turn out to be \emph{distinct} theories. Basically, the problem of formulating a dynamics for quantum entities---whatever they are---is now known as the \textit{measurement problem}.

According to the standard way of formulating the measurement problem (Maudlin \cite[p. 7]{mau95}), the following three apparently reasonable propositions are inconsistent (despite being pairwise consistent):
\begin{quote}
\begin{itemize}
    \item[1.A] The wave-function of a system is \textit{complete}, i.e., the wave-function specifies (directly or indirectly) all of the physical properties of a system.
    \item[2.A] The wave function always evolves in accord with a linear dynamical equation (e.g., the Schrödinger equation).
    \item[3.A] Measurements of the spin of an electron (typically) have determinate outcomes, i.e., at the end of the measurement, the measuring device is either in a state that indicates spin up (and not down) or spin down (and not up). (Maudlin \cite[p. 7]{mau95})
\end{itemize}
\end{quote}

The major difficulty to ensure consistency comes from having the Schrödinger equation (or something to the same effect) as the dynamics of quantum systems and having to account for specific and determinate measurement results by using that very dynamics. The problem derives from the linearity of the equation and the probabilistic character of quantum theory. When one attempts to describe a measurement situation composed by a quantum system plugged into a measurement apparatus through such an equation, one ends up with superpositions of distinct states of the system being measured and distinct measurement results. The entire system (quantum system + measuring apparatus) typically does not evolve, following the Schrödinger equation, to a definite eigenstate of the observable being measured and the measuring apparatus pointing to the correspondent result. Rather, the whole system evolves in a state of superposition of different states of the system and different states of the apparatus (associated with distinct measurement results). However, such superpositions are never observed: only determinate results are. Hence, to restore consistency, something needs to go.

To provide a solution to the measurement problem is to offer an interpretation of quantum theory. Some interpretations deny that the wave function is complete (hidden variables interpretations); others challenge that the Schrödinger equation is the only account of the dynamics; yet others question that we need determinate outcomes (non-collapse interpretations).

Relevant to us is that the measurement problem is closely connected to the problem of providing a dynamics to quantum theory. However one addresses this issue, a problem remains. As noted above, following Esfeld, a physical theory must provide not only the dynamics, but also an ontology, that is, a more or less explicit characterization of the entities that are posited by the theory. How is the problem of ontology related to the measurement problem?

The two problems are closely connected. A dynamics is a dynamics for the entities postulated by the theory. Different interpretations populate the world with distinct kinds of entities. The many-worlds interpretation posits a plurality of worlds, whereas Bohmian mechanics puts forward particles endowed with trajectories and a pilot wave. Attempts to explain measurements as acts of conscious activity, in turn, postulate the existence of consciousness apart from matter. Clearly, interpretations contribute directly to an ontology. As Laura Ruetsche \cite[p. 3433]{rue15} notes, when one believes in a theory, one believes in an interpretation of the theory, which provides an account of what the world is like according to it. Interpretations articulate ontologies for theories.

If ontology is the part of metaphysics concerned with determining the world's furniture, to solve the measurement problem involves describing the ontology of quantum mechanics. In this respect, other chapters in this book advance distinct ontologies as part of their account of the measurement problem. This makes ontology importantly related to scientific development, and not something of just a philosophical concern, making room for a partially naturalistic approach to ontology (Esfeld \cite{esf19}, Sklar \cite{skl10}).

In addition to the particular entities postulated by each interpretation, there is much in common among various interpretations. Most of them posit the existence of atoms, electrons, protons, etc. (Ladyman \cite{lad19}). These entities are also put forward in each formulation of quantum theory; hence, they are part of the theory's ontology. But this is not the case for every interpretation of quantum mechanics. Wave function realism, for instance, populate the world with a wave function, from which everything else (including particles) should follow. However, even if these entities are a byproduct of the fundamental ontology, they must be recovered somehow, and the question regarding their metaphysical status is still pertinent. A major problem concerns specifying precisely what kind of things the entities that are typically dealt with by quantum mechanics are: particles or fields? What is the picture of reality suggested by the theory?

Modern physics---including quantum mechanics---is formed, to a large extent, by field theories. Quantum physics is significantly concerned with the study of \textit{quantum fields}. ``Particles'' arise as second-order entities, emerging when fields interact and momentum and energy are transferred, so that certain field excitations arise. The excitations (the ``particles'')  come in units, namely, in $1, 2$, \ldots, $n$ units of excitation. The interactions increase or decrease the number of excitations by one and they are supposed to happen in space and time, so that the excitations behave as particles, hence the term. However, when an experiment with cathode rays is conducted, one (supposedly) see \textit{electron} beams, rather than electrons, despite the way one describes them. When neutral hydrogen atoms are ionized, it is with electrons that one interacts empirically, not with a mathematical field. When something is accelerated, mathematical fields or wave functions are not accelerated, but ``particles'' are. Although the mathematical formalism refers to a mathematical entity called `field', a particle interpretation needs to be provided to accommodate with these empirical circumstances. The vocabulary of experimental physics is mostly a vocabulary of particles. Even if we can dispense with the notion of particle when it comes to mathematics, talk of particles still remains in the experimental front. Michael Redhead talks about the ``particle grin'', by comparing the situation with the Cheshire cat (French and Krause \cite[p. 361]{fre06}).

In orthodox quantum mechanics, before the advent of quantum field theories, the systems were supposed to behave either as particles or as waves, but never as both. The terms ``particles'' and ``waves'' were taken from analogies with what was known from classical physics and ordinary experience. In quantum mechanics, there is limited knowledge of the \textit{real} entities under investigation, except for what is postulated by the theory.

These are the \textit{quantum objects} we will focus on. According to the Received View of quantum non-individuality, the behavior of entities of this kind, independently of whether they are described by orthodox quantum mechanics or field theories, is not compatible with the traditional metaphysical conception of \textit{individuals}. This chapter aims to make these points clear. But, first, we need to say a bit more about the key concepts involved in the metaphysics of individuality.

\section{Non-individuality and metaphysical profiles}\label{sec3}

Once it is acknowledged that particles of some kind need to be accommodated, a further question arises: what kind of things are these objects? A philosophical theory about objects may be considered, and the question about whether quantum entities are individuals becomes central. Moreover, if the tradition is followed, and if particular entities must have an individuation principle, according to which principle of individuation should we individuate quantum entities?

To come to grips with these questions, a proper understanding of principles of individuation is called for. We will consider the issue in the next section. First, we motivate the need for engaging with metaphysics when it comes to quantum entities. Historically, such entities have been treated as lacking individuality, in an informal sense. Some of the founding fathers of quantum theory noted the odd behavior of quantum entities, and thought that this required close consideration (French and Krause \cite[Chapter 3]{fre06}). Erwin Schrödinger, in particular, examined the issue attentively. He starts an essay on the nature of quanta explicitly claiming that quantum entities have no individuality:

\begin{quote}
This essay deals with the elementary particle, more particularly with a certain feature that this concept has acquired---or rather lost---in quantum mechanics. I mean this: that the elementary particle is not an individual; it cannot be identified, it lacks ``sameness''. [\dots] In technical language it is covered by saying that the particles ``obey'' a new fangled statistics, either Bose-Einstein or Fermi-Dirac statistics. The implication, far from obvious, is that the unsuspected epithet ``this'' is not quite properly applicable to, say, an electron, except with caution, in a restricted sense, and sometimes not at all. (Schrödinger \cite[p. 197]{sch98})
\end{quote}

Note that Schrödinger directly relates what he saw as a new metaphysical character of quanta (they are not individuals, cannot be identified, lack ``sameness'') with quantum statistics. Quantum statistics is connected with permutation symmetry in quantum mechanics, which leads to quanta being indiscernible by any observable available in the theory. As will become clear, this relation between statistics and new metaphysical features of the entities is not as direct as Schrödinger expected, although any kind of metaphysical approach to quantum mechanics will need to accommodate the odd statistics and be compatible with it.

Crucial for us is that the lack of individuality, which emerges directly from quantum theory (at least from a historical point of view), is a metaphysical question about the metaphysical status of the particles regarding their individuality. While some may think that considering the nature of quantum entities may lead us too far from the theory itself---due to the fact that the problem of individuality is a metaphysics problem---Schrödinger (among others) saw the connection between non-individuality and physics as a pretty direct one. Furthermore, according to some philosophers, it is only after such questions about the nature of the relevant entities are properly addressed that one can be confident of having a \textit{clear picture} of what one is accepting when the theory is accepted. Steven French, in particular, dubs this demand for a metaphysical picture \textit{Chakravartty's challenge}, given Anjan Chakravartty's identification of the need for a metaphysical clarification for a reasonable articulation of realism (French \cite{fre14}; French and Krause \cite[p. 244]{fre06} explicitly emphasize the need for a metaphysical articulation of the ontology). Whether Chakravartty himself agrees with this understanding of his own work is a matter of controversy (Chakravartty \cite{cha19}). French continues:

\begin{quote}
But how do we obtain this clear picture? A simple answer would be, through physics which gives us a certain picture of the world as including particles, for example. But is this clear enough? Consider the further, but apparently obvious, question, are this particles individual objects, like chairs, tables, or people are? In answering this question, we need to supply, I maintain, or at least allude to, an appropriate metaphysics of individuality, and this exemplifies the general claim that in order to obtain Chakravartty's clear picture and hence obtain an appropriate realist understanding we need to provide an appropriate metaphysics. Those who reject any such need are either closet empiricists or `ersatz' realists. (French \cite[p. 48]{fre14})
\end{quote}

In other words, in addition to the ontology---the items posited by a theory---further metaphysical questions still need to be addressed (Arenhart \cite{are19}). In particular, if the entities a theory posits are particulars, questions about identity and individuality arise quite directly.

This suggests that a further interpretational layer must be added to the posits of a theory, a layer involving something like the metaphysical profile of the relevant entities. This metaphysical layer does not perform any role in the solution to the measurement problem, or to fix the ontology that comes with such solution. Rather, it is presupposed that the ontology is already in place, and one works from there. This suggests that one cannot \textit{extract} the metaphysical profile of the entities from the theory, as Schrödinger had suggested, but rather must impose it from above.

One could ask why this extra step is required. As French suggested in the quotation above, addressing these issues is the only way to articulate a completely realistic view of what the world is like according to the theory in question. Otherwise, something will be missing. But whether or not one takes this extra step, we take it that individuality and non-individuality are metaphysical issues that add an additional layer to the description of reality provided by quantum theory. In this sense, one shifts from a mere operational use of quantum theory to an interpreted version of the theory, and from there, to an interpretation plugged with a metaphysical characterization of its posits. This leads one, in principle, to a fuller version of the theory.

We can now examine what individuality is (and its absence), and some of the accompanying concepts of identity, indiscernibility, and identification.

\section{Individuals, individuation, and identity}\label{sec4}

There is a lot of ambiguity in discussions of quantum identity and individuality. Unfortunately, such ambiguities contaminate also formulations of the Received View and, as a consequence, some of its criticisms. In this section, we specify more precisely the meaning of the terms we use. This will inform our examination of the approaches to the Received View that will be developed in the next sections (see also Krause and Arenhart \cite{kra18}, and Arenhart, Krause and Bueno \cite{are19a}).

By \emph{identity} we mean the \textit{logical relation} of identity, typically symbolized by the binary predicate symbol `='. A sentence involving the identity symbol, such as `$a = b$' describes a relation that, intuitively speaking, is true when $a$ and $b$ are names of one and the same thing, and is false when $a$ names a distinct object from $b$. Obviously, this explanation of the meaning of identity is circular, but it is not supposed to work as a definition of identity. (The issue of whether identity can be defined or explained in more basic terms is delicate, and many argue that identity is fundamental; see Bueno \cite{bue14} and Shumener \cite{shu17} for further discussion.)

Typically, identity is presented by its first-order logical axioms:

\begin{description}
\item[Reflexivity] $\forall x(x = x)$
\item[Substitution law] $x = y \rightarrow (\alpha \rightarrow \alpha[y/x])$, with the known restrictions.
\end{description}

\noindent The law of substitution, in its first-order version, states that terms that are related by identity in any formula can be replaced by one another without disturbing the truth value of such formula. In material terms, it becomes the law of the \textit {indiscernibility of identicals}, which states that identical things are indiscernible by properties and relations. With regard to the relation between identity and indiscernibility, this is only half of the story, with identity implying indiscernibility. The other half is much more controversial, stating that indiscernibility implies identity. This is the famous \textit{principle of identity of indiscernibles} (PII). In material terms, it states that items that share all properties and relations are numerically identical. In first-order languages, PII is stated as a schema:

\begin{description}
\item[PII] $(\alpha(x) \leftrightarrow \alpha(x/y)) \rightarrow x = y$
\end{description}

PII has counterexamples in first-order languages (it is enough to use a poor language), not being a truth of first-order logic. The metaphysical thesis (the material reading), however, has proved more resistant. In quantum mechanics, controversy around the principle is notorious. While some have claimed that it is a false principle in quantum theory (French and Krause \cite[Chapter 4]{fre06}), others have tried to save weaker versions of the principle, going as far as to claim that quantum mechanics proves weaker versions of PII (Muller and Saunders \cite{mul08}). We will not engage with this particular controversy in this chapter.

Finally, we need also to specify what is meant by the notion of \textit{individual}. According to the \textit{Online Etymology Dictionary}, ``individual'' is a Medieval term that means ``single object or thing''. In other usages of the term, for instance in Middle English, ``\textit{individuum} was used in [the] sense of `individual member of a species' (early 15c.)''. In current metaphysics, one of the tasks of metaphysicians consists in explaining what is it that confers individuality to an individual, that is, what is it that makes that individual precisely the individual that it is. As Jonathan Lowe \cite[p. 75]{low03} has put it, the role of an individuation principle with regard to an object is to account for ``whatever it is that makes it the single object that it is---whatever it is that makes it \ita{one} object, distinct from others, and the very object that it is as opposed to any other thing''.

The notion of individual is related to the notion of numerical identity, but not, necessarily, with the notion of discernibility. In other words, individuality does not exclude the possibility that ``different'' individuals have the same qualitative features. After all, an ``individual member of a species'' need not be \textit{discernible} from other members. The additional supposition that every individual should be a unity \textit{discernible} from other individuals was a central piece in Leibniz's metaphysics, which adopts PII. This principle, in turn, is responsible for the connection between identity and indiscernibility. The idea that identity is grounded in qualitative features has its epistemic attractions, but it is not required by the notion of individuality. It remains an open issue whether to adopt an individuality principle that links identity with indiscernibility as articulated by PII, or else to leave open the possibility that individuals can be indiscernible.

\section{Non-individuals, lack of identity}\label{sec5}

As just noted, the concepts of identity, individuality and indiscernibility can be understood as being largely independent. These concepts can now be used to frame a metaphysics of non-individuality; one that contributes to the metaphysical profile of quantum entities. This is required for a more complete picture of what the world looks like according to quantum mechanics.

Schrödinger, it was noted above, declared that quantum entities lost their individuality. What emerges from this, we take it, is the need to specify a connection between quantum theory and a metaphysics of non-individuality. To examine this issue, consider another passage in which Schrödinger stresses the point:

\begin{quote}
I beg to emphasize this and I beg you to believe it: it is not a question of our being able to ascertain the identity in some instances and not being able to do so in others. It is beyond doubt that the question of `sameness', of identity, really and truly has no meaning. (Schrödinger \cite[pp. 121-122]{sch96})
\end{quote}

\noindent Strictly speaking, Schrödinger is claiming that sameness and identity fail to make sense for quantum entities. This is not obviously the same as to claim that quantum entities are non-individuals in light of the distinctions we made above. How can we make sense of this claim metaphysically? In what follows, a particular approach is developed that puts some metaphysical flesh on the bones of the Received View, as described above, and which attempts to capture the literal reading of `losing identity', as Schrödinger suggested.

Consider again the notion of individuality. There must be something that accounts for what an entity is, and makes it precisely that entity. Some have thought that an entity must be an individual in virtue of possessing a special non-qualitative property of being identical to itself, and that this is what explains its individuality. Socrates is an individual in virtue of being endowed with the property of `being identical to Socrates'. Nothing else has this property, only Socrates. Hence, this accounts for the difference between Socrates and everything else, as a kind of individual essence. At the same time, this is a non-qualitative property: one cannot use it to \emph{qualitatively} distinguish Socrates from anything else. That means that, in terms of qualities, two individuals may coincide, while still having different essences---their non-qualitative individuating properties. In terms of the concepts formulated in the previous section, two items may be numerically distinct individuals, but at the same time be qualitatively indiscernible. This non-qualitative property is called in the literature a `haecceity'. Individuals are the individuals they are in virtue of their haecceity. Typically, a haecceity is expressed in terms of self-identity: Socrates's haecceity is expressed by the fact that `Socrates = Socrates'. It is, in other terms, the property of `being identical to Socrates'.

This approach to individuality accommodates the link between individuality and identity, and also, as French and Krause \cite{fre06} have observed, between lack of identity and lack of individuality, which is presumably what Schrödinger had in mind---although Schrödinger would not have made the point in terms of haecceity. French and Krause consider principles of individuality that rely on non-qualitative features to provide a form of transcendental individuality---and haecceitism is not the only one available, although it is our focus here (French and Krause \cite[Chapter 1]{fre06}). If individuality is related to identity as just suggested, failure of identity amounts to failure of individuality. The case for non-individuality is then made metaphysically: a non-individual is something that has no haecceity, and given what haecceities are, non-individuals are entities for which self-identity fails. As French and Krause note:

\begin{quote}
[\dots] the idea is apparently simple: regarded in haecceistic terms, ``Transcendental Individuality'' can be understood as the identity of an object with itself; that is, `$a = a$'. We shall then defend the claim that the notion of non-individuality can be captured in the quantum context by formal systems in which self-identity is not always well-defined, so that the reflexive law of identity, namely, $\forall x(x = x)$, is not valid in general. (French and Krause \cite[pp. 13-14]{fre06})
\end{quote}

We will sketch below a way of articulating the approach formally, preserving the close connection between identity and individuality that is central to the conception. As French and Krause point out:

\begin{quote}
We are supposing a strong relationship between individuality and identity [\dots] for we have characterized `non-individuals' as those entities for which the relation of self-identity $a=a$ does not make sense. (French and Krause \cite[p. 248]{fre06})
\end{quote}

As a result, it is possible to advance a metaphysical conception of non-individuality along the lines suggested by Schrödinger. Understood in this way, non-individuals are indiscernible by their properties. This is a physical issue, rather than a logical or philosophical one. We may approach indiscernibility in the way physicists do: the relevant entities are characterized by physical laws. Electrons, for instance, are those entities specified by a law according to which a mass $m = 9.109 \times 10^{-31}$ kg and a charge $e = −4.80320451(10) \times 10^{-10}$ esu is always accompanied by spin $1/2 \hbar$, a Bohr magneton, etc. (see Toraldo di Francia \cite[p. 342]{tor81}; `esu' is the abbreviation for the electrostactic unity of charge). These entities are, on di Francia's suggestive terminology, \textit{nomological} objects. In other words, one does not analyze these objects in order to obtain their properties; the properties are specified by the relevant theory. It then emerges that all electrons are absolutely indiscernible. (This idea can be questioned, though; see Dieks \cite{die11}.)

\subsection{The mathematics}

On the Received View, we noted, non-individuals are entities devoid of a haecceity and, hence, cannot figure in the relation of identity. They are also nomological entities, which are specified by physical laws, and do not obey the standard theory of identity. We will examine it now.

The considerations so far motivate the development of a formal system to accommodate non-individuals, along the lines suggested by French and Krause \cite{fre06} (see, in particular, the quote above). The key idea is that these entities can be aggregated in collections, which are called \textit{quasi-sets} and can have a cardinal, but not an ordinal. Non-individuals do not bear unambiguous names or identifications. Given the lack of a haecceity, all of them would have the same ``identity card'', which is, thus, useless. But non-individuals of different kinds (e.g. electrons and positrons) can be distinguished from one another, such as, by their electric charge (the positron's charge is the opposite of the electron's). However, since they do not obey the theory of identity, it is not appropriate to claim that they are \textit{different}. Talk of \textit{discernible} and \textit{indiscernible} is enough.

All of this conflicts with standard logic and mathematics. Let us start with the notion of identity. Identity can be either defined or taken as primitive. Suppose we have a language with just three primitive predicate symbols: $P$ and $Q$ are unary symbols, whereas $R$ is binary. It can then be stated that:
\begin{equation}
\begin{aligned}
    x = y := \; & (P(x) \lra P(y)) \wedge (Q(x) \lra Q(y)) \\            & \wedge
    \forall z (R(x,z) \lra R(y,z))
             \wedge \forall z (R(z,x) \lra R(z,y)).
\end{aligned}
\end{equation}

This can be generalized to any finite quantity of predicates. Of course, this offers no definition of numerical identity, and provides only indiscernibility relative to the language's predicates. It would be better to indicate this fact with $x \sim y$. The same objection can be raised against any attempt to use equivalence relations or congruences to stand for identity. For reasons that will emerge shortly, these strategies only hide the problem, making individuals appear to be non-individuals, while they are not. But why are they not?

The reasons have to do with set theory, the typical framework in which to develop all standard mathematics. Take a system such as ZFC, Zermelo-Fraenkel set theory with the Axiom of Choice (we are not including ur-elements for now). Axiomatized as a first-order theory, as is now usual, ZFC has `$=$' as a primitive binary symbol, satisfying the following axiom schema:

\begin{enumerate}
    \item ($=_1$) $\forall x (x=x)$ (Reflexive Law of Identity).
    \item ($=_2$) $\forall x \forall y (x = y \to (\alpha(x) \to \alpha(y)))$, in which $\alpha(x)$ is a formula (that may contain parameters), $x$ is free and $\alpha(y)$ is obtained from $\alpha(x)$ by the substitution of some free occurrences of $x$ by $y$ (Axiom of Substitution of Identicals).
    \item ($=_3$) $\forall x \forall y (\forall z (z\in x \lra z \in y) \to x = y)$ (Extensionality Axiom).
\end{enumerate}

The converse of the extensionality axiom follows from ($=_2$). It is also easy to prove that identity is symmetric and transitive. For our purposes, the key consequence is that the universe of sets---the von Neumann cumulative hierarchy of well-founded sets $\mcal{V} = \langle V, \in \rangle$ (Jech \cite{jec03})---does not have nontrivial automorphisms. This has important implications.

First, from a set-theoretic perspective, mathematics is conducted inside a mathematical structure. A group is a structure of the kind $\mcal{G} = \langle G, \star \rangle$; a vector space is a structure of the form $\mcal{V} = \langle V, K, +, \cdot \rangle$; a classical particle mechanics is a structure $\mcal{CPM} = \langle P, \mbf{s}, m, \mbf{f}, \mbf{g} \rangle$ (McKinsey et al. 1953), and so on. These structures are (or can be viewed as) \textit{sets} of ZFC. Let $\mcal{E} = \langle D, R_i \rangle$ be a structure, in which relations $R_i$ hold for the elements of the domain $D$ (all other structures can be reduced to this case). Two objects $a$ and $b$, which belong to $D$, are $\mcal{E}$-indiscernible if there is an automorphism $h$ of $\mcal{E}$ such that $h(a) = b$ (thus, $h^{-1}(b) = a$). As an illustration, $i$ and $-i$ are $\mcal{C}$-indiscernible in the complex field structure $\mcal{C}$, for the operation of taking the conjugate is an automorphism. Structures that have automorphisms other than the identity function are said to be \ita{deformable}, or \ita{non-rigid}; otherwise, they are \ita{rigid}.

A significant theorem states that \textit{every} structure can be extended to a rigid structure (da Costa \& Rodrigues \cite{cos07}). This means that the apparently indiscernible objects of the domain of a given structure can be individuals with identity on the outside of it. Furthermore, recall that the whole universe is rigid. So, within a mathematical framework, such as ZF, ZFC, NBG, KM and other ``standard'' set theories, indiscernible things can only be taken to be $\mcal{E}$-indiscernible for some structure $\mcal{E}$. This maneuver is adopted when congruence relations are used to stand for identity. Good foundational work should avoid this trick.

To conclude this section, some remarks on set theories with ur-elements are in order (see Suppes \cite{sup60}). Ur-elements are entities that are not sets but can be members of sets. They do not have elements, relative to a membership relation, although they can obey mereological axioms. Set theories with ur-elements are useful for applications in the empirical sciences, in which many objects, such as eucalyptuses, elephants, and electrons, are not sets, but can be members of them.

Let $A$ be the set of ur-elements. It is known that every permutation on $A$ can be extended to an automorphism of the entire universe, which is not rigid, in contrast to the universe of ``pure'' sets. This means that, in certain situations, two ur-elements apparently could not be discerned by any means and so they might be good candidates to represent quantum objects. However, this should be viewed with suspicion. Even ur-elements obey the axiom of pairing, which states that given sets $a$ and $b$ of ur-elements, there is a set $z$ that has only $a$ and $b$ as its elements. If $a=b$, the unitary of $a$ is denoted by $\{a\}$. The ``property'' $I_a(x) \lra x \in \{a\}$, which is called \ita{the identity of $a$} and is satisfied only by $a$, can then be defined. This property can be used to establish, as a consequence of PII, that $a$ has something that is not shared by any other object in the universe. Hence, although ur-elements can be taken to be indiscernible, they are not. They do have identity: every ur-element is \textit{distinct} from any other entity in the universe.

\subsection{A more appropriate mathematics}

We noted that if quantum objects are non-individuals, this does not entail that they cannot be discerned in some cases: quantum entities of different kinds, such as electrons and protons, can be distinguished. But the fact that these objects are discernible in certain instances does not grant them identity. Since standard set theories presuppose the identity of all objects in their domains, they are not adequate to study entities of this sort. To address this issue, a novel set theory has been articulated. We now provide some details about it.

Quasi-sets are collections of objects some of which lack identity conditions, that is, they are non-individuals. Other elements satisfy the rules of standard logic, including those of identity. Those that do are \textit{individuals}; they are either sets or ur-elements. When restricted to individuals, the theory yields a copy of ZFU, Zermelo-Fraenkel set theory with ur-elements. The latter are $M$-objects, whereas non-individuals are $m$-objects ($M$ for ``macro'', $m$ for ``micro''). The theory allows for $m$-objects of several kinds, and can accommodate protons, electrons, positrons, and so on. Although $m$-objects of a given kind, e.g. electrons, may or may not be discerned, one cannot claim that they are \textit{different}, for identity does not apply to them.

Operations on quasi-sets are similar to those on sets: intersection, union, and power set, for instance, are all specified for quasi-sets. A quasi-set may have a cardinal, but when indistinguishable $m$-objects are involved, no ordinal can be associated to it. So, a quasi-set cannot be ordered, named, or labelled without ambiguity (see French and Krause \cite{fre06}). Given these remarks, we can now examine how this theory can be used to do physics.

\subsection{Quantum physics}
The main philosophical problem regarding quantum physics does not concern the results of the theory. These are well established and properly supported. The problem concerns interpretation. Our aim is to present a reasonable account of quantum theory that takes quantum objects, which are, in any case, involved in any interpretation, as non-individuals. We consider how a different mathematical theory can accommodate this point.

The application of quasi-set theory involves various steps, culminating in a formulation of quantum theory in terms of quasi-sets. On standard formulations, quantum objects themselves (or quantum systems) appear only in an indirect way. Despite Max Jammer's comment to the effect that the theory's primitive notions include  \textit{system} (Jammer \cite[p. 5]{jam74}), the standard formalism is formulated in terms of \textit{states} and \textit{observables}. (There is no need to review the axioms here. We just note that there are plenty of different alternative ways of formulating quantum mechanics: Styer et al. \cite{st02} present nine.) On the standard formulations, it is assumed that there is a \textit{set} of quantum systems to begin with. However, \textit{sets} from classical set theories, such as those mentioned above, are collections of \textit{distinct} objects. The members of a set must have identity: they are individuals. Thus, something goes wrong if sets are assumed in this context. Our proposal is to assume that quantum systems form a quasi-set instead.

At least two ways are available to accommodate quantum mechanics in quasi-set theory. The first concerns non-relativistic quantum mechanics, QM. The core idea was advanced in Krause and Arenhart (2017, \S 5.8.1), and explicitly assumes, as in the case of classical particle mechanics mentioned earlier, that there is a quasi-set of quantum systems, $S$, such that some of them, depending on the particular model under consideration, can be indiscernible. The structure has the form:

\begin{equation}
    \mcal{QM} = \Big\langle S, \{\mcal{H}_i\}, \{\mcal{\hat{A}}_{ij}\}, \{\mcal{U}_{ik}\},  \mathcal{B}(\mathbb{R})\Big\rangle, \; i \in I, j \in J, k \in K
\end{equation}

\noindent in which:
\begin{enumerate}
\item $S$ is a quasi-set whose elements are called \ita{physical objects}, or \ita{physical systems}.
\item $\{H_i\}$ is a family of complex separable Hilbert spaces whose cardinality is defined in the particular application of the theory.
\item $\{\hat{A}_{ij}\}$, for each Hilbert space $\mcal{H}_i$, is a family of self-adjunct (or Hermitian) operators over $H_i$.
\item $\{U_{ik}\}$, again for each $\mcal{H}_i$, is a family of unitary operators over $H_i$.
\item $\mathcal{B}(\mathbb{R})$ is the collection of Borel sets over the set of real numbers.
\end{enumerate}

The Hilbert space formalism, whose postulates we will review in a moment, does not involve space and time, something needed in order to apply it to the world. We will return to this point.

To each quantum system $s \in S$, we associate a 4-tuple of the form:

\begin{equation}
\sigma = \langle \mathbb{E}^4, \psi(\mathbf{x}, t), \Delta, P\rangle,
\end{equation}

\noindent in which $\mathbb{E}^4$ is the Galilean spacetime (see Penrose \cite[Chapter 17]{pen07}), such that each point is denoted by a 4-tuple $\langle x, y, z, t\rangle$, in which $\mathbf{x} = \langle x, y, z \rangle$ denotes the coordinates of the system and $t$ is a parameter representing time. $\psi(\mathbf{x},t)$ is a function over $\mathbb{E}^4$, which is called the \ita{wave function} of the system. $\Delta \in \mathcal{B}(\mathbb{R})$ is Borelian. $P$ is a function defined in $\mathcal{H}_i \times \{\hat{A}_{ij}\} \times \mathcal{B}(\mathbb{R})$, for some $i$ (determined by the physical system $s$), and having values in $[0,1]$. Thus, $P(\psi, \hat{A},  \Delta) \in [0,1]$ is the probability that the measurement of the observable $A$ (represented by the self-adjunct operator $\hat{A}$) for the system in state $\psi(\mathbf{x},t)$ lies in $\Delta$.

The relation between the state vector and the wave function is given as follows: Let $(\mbf{x},t)$ denote the position operator at time $t$. Then $\psi(\mbf{x},t) = \langle (\mbf{x},t) | \psi\rangle$, that is, the wave-function is described by the Fourier coefficients of the expansion of the state vector in the orthonormal basis of the position operator.

Suitable postulates connecting all these concepts need not be reviewed here (Krause and Arenhart 2017). The key shift is the adoption of quasi-set theory. This approach is able to resolve the difficulties raised by Dalla Chiara and Toraldo di Francia \cite{dal93} about the interpretation of a language $L$ of quantum mechanics if it is formulated in a standard set theory, such as ZFC. We review some of their arguments and explain why these criticisms do not hold in quasi-set theory.

\begin{enumerate}
    \item The standard interpretation assumes that its domain $D$ is a set of individuals. In contrast, in quasi-set theory, the domain includes a quasi-set of indiscernible entities, so that some quantum objects can be taken to be indiscernible without the usual tricks.
    \item For any individual $d \in D$ in the standard interpretation, the language can be extended to a language $L'$, which contains a name $a$ and a denotation function $\rho$, such that $\rho(a) = d$. In quasi-set theory, this is not the case, for $m$-objects cannot be named. In fact, a suitable language for QM should not allow for proper names.
    \item If $L$ is at least a second-order language, Leibniz's PII holds in standard semantics:
    $$\forall x \forall y (x \not= y \to \exists F (F(x) \wedge \neg F(y))).$$ In contrast, in quasi-set theory, this principle fails for indiscernible $m$-objects, for no haecceity is assumed and they may share all their properties and relations.
\end{enumerate}

The second approach to use quasi-set theory to accommodate quantum mechanics focuses on the Fock space formalism. This was started in Domenech et. al. \cite{dom08} and \cite{dom10}. The Fock space formalism includes creation and annihilation operators, making room for (non-interacting) fields. In the papers just mentioned, two kinds of vector spaces (Hilbert spaces) are defined; they are called \textit{quasi-}spaces---one for bosons, another for fermions. The main idea is that, by positing that some entities can be strongly indiscernible, it is possible to dispense with certain postulates, such as the indistinguishability postulate. Quantum results emerge naturally with this move.

Moreover, within quasi-set theory, the so-called quantum statistics (B-E and F-D) are obtained quite naturally too. Given that certain objects can be indiscernible in a strong sense, the formulas that provide the counting states in both B-E and F-D statistics can be obtained just by counting the states---without having to discharge some of them, as is usual in the standard approach, to mimic indiscernibility. In this respect, the quasi-set-theoretic framework seems to be much more natural for the quantum domain (French and Krause 2006, \S 7.6).

In quasi-set theory, not only indiscernible \textit{objects} can be considered, but also indiscernible \textit{properties}. As is well known, to arrive at statistical predictions from the quantum formalism, the same experiments need to be repeated enough times to allow for the computation of mean values and probabilities, and the determination of all necessary stochastic properties of experimental outcomes. But what counts as ``the same experiment''? De Barros, Holik and Krause \cite{bhk19} elaborated on this notion to underscore how indistinguishability is deeply connected to contextuality. They realized that in order to perform $n$ experiments, physicists must prepare $n$ ``identical'' copies of a same quantum system. But, given the indistinguishability postulate---according to which different mean values in permutations of indistinguishable objects is not observable---the particles involved cannot be identified. Interestingly, this is so even if we perform a thought experiment, provided that we want to respect what the logic of QM seems to suggest regarding identity, at least given an interpretation of quantum theory.

When considering indistinguishable properties, we neither perform ``the same'' experiment twice nor measure two indistinguishable properties on ``the same'' quantum system, but we measure indistinguishable properties---prepared in ``the same'' way---over indistinguishable quantum systems. Thus, by seriously considering the notion of indistinguishability as something distinct from identity, something that quasi-set theory enables us to do, we can read the results by Kochen and Specker and realize that the core of their theorem (the ``paradox'') can be avoided, for the contradiction assumes that ``the same'' properties are measured in ``the same'' particles in different contexts. However, if we assume the apparently obvious fact that we measure indistinguishable properties over indistinguishable particles, there will be no surprise in acknowledging that the obtained results may differ. In other words, within this context, the Kochen-Specker result can be put within parentheses (de Barros et al. \cite{bhk19}).

\section{Non-individuals with identity}\label{sec6}

Despite the developments in the Received View, some may argue that the loss of identity, as an expressive resource, is just too much to swallow. The relation between identity and individuality that was suggested leads to a relation of identity heavily burdened by the metaphysical role of individuation---a role that, according to some, identity should not have. But for those who want to keep the idea that quantum mechanics does not deal with individuals, some options are still available. Given the distinctions we provided regarding identity and individuality, loss of individuality need not entail loss of identity. So, even if one is not willing to embrace the version of the Received View above, there is no need to abandon the Received View altogether, formulated as the thesis that quantum entities are non-individuals. It is enough to apply distinct principles of individuality, and to consider those whose failure do not imply lack of identity (Arenhart \cite{are17}, and Krause and Arenhart \cite{kra18}).

Here we present an alternative version of the Received View that allows us to have non-individuals while identity is still maintained. Of course, in this context, identity and individuality must be separated: the fact that one may use identity to refer to some entities does not imply that they have a haecceity. To do so, identity should be deflated from the metaphysical content it had been burdened with in the previous approach, and individuality endowed with more epistemic features. Recall that an individual is a unity of some kind. The idea gains a twist with the addition that an individual is something that can be repeatedly identified as being \textit{the same} individual, that is, it requires identification over time.

Bueno's approach to individuality in \cite{bue14} provides one such option. According to his approach, an individual must satisfy two conditions:

\begin{description}
\item[Discernibility:] The item is discernible from any other individual.
\item[Re-identification:] The item is re-identifiable over time.
\end{description}

\noindent Although identity can be employed in the formulation of these conditions, the fact that certain entities have identity, by itself, does not confer them individuality. As Bueno \cite{bue14} remarks, this account of individuality is compatible with a view in which the relation of identity is devoid of metaphysical content. Interestingly, a notion of non-individuality also emerges: a non-individual is something that fails to be an individual. As a result, three kinds of non-individuals are obtained. Something is a non-individual provided that (i) it is indiscernible from others of its kind; or (ii) it lacks conditions of identification over time; or (iii) it is indiscernible from other entities of its kind and lacks conditions of identification over time. Quantum entities, on a reasonable interpretation, are non-individuals according to the third option.

As noted above, the typical version of the Received View states that non-individuals are entities without identity. However, if the Received View is merely the thesis that quantum entities are non-individuals, the proposal we have presented here also provides a formulation of the Received View. This approach has an interesting historical motivation. Consider again Schrödinger's quote that supports the previous version of the Received View:

\begin{quote}
I beg to emphasize this and I beg you to believe it: it is not a question of our being able to ascertain the identity in some instances and not being able to do so in others. It is beyond doubt that the question of `sameness', of identity, really and truly has no meaning. (Schrödinger \cite[pp. 121-122]{sch96})
\end{quote}

In this passage, Schrödinger is not completely clear about what it is meant by `identity'. Perhaps the failure is not of numerical identity, but rather of identity over time. As he notes, in a given experiment, one cannot claim, with full certainty, that a particle that appears at an instant $t_1$ in a bubble chamber is the same that appears at a later instant $t_2$. The particle's identity cannot be determined because it cannot be re-identified in distinct instants of time. So, what fails is not the particle's numerical identity, but its identity over time. This makes such a particle a non-individual according to proposal (ii) above.

If non-individuals are items lacking individuality in this second sense---that is, they cannot be re-identified over time---Hans Dehmelt's positron Priscilla as well as other quantum objects he has trapped are non-individuals. After all, as soon as they leave the apparatus that trapped them, these objects cannot be identified anymore. This is not just an epistemological feature, but belongs to the very nature of these entities. (The same happens with more recent trapped quantum objects by Haroche and Wineland; for a discussion of Priscilla, see Krause \cite{kra11}.)

One of the benefits of this approach is that it does not require revising the logic of identity. Quantum theory, however, does not favor one metaphysics over another. Thus, not only is quantum mechanics empirically underdetermined, given that its data are compatible with distinct interpretations, but it is also metaphysically underdetermined: there is no specific quantum mechanical support for a preferred metaphysical interpretation of its entities. This is as it should be: metaphysical interpretations were designed to provide answers to metaphysical questions rather than to problems connected to the applications of the theory.

\section{Conclusions}\label{sec7}
In quantum mechanics, similarly to what happens in arithmetic, one starts with informal notions. A puzzle emerges when it is realized that quantum ``particles'' have lost their individuality. To make sense of this, a more fine-grained framework, which does not presuppose the identity of the objects under consideration, is called for. Such framework, articulated in quasi-set theory, offers the basis for an interpretation of quantum mechanics with regard to the nature of the entities in question.

Of course, one can be an instrumentalist about quantum theory, refuse to engage with these issues, and insist that QM only provides tools for computing probabilities. But something would be missing. One can be legitimately interested in the world views that QM offers. One of them suggests that quantum objects are better thought of as non-individuals. We take this view seriously, and have provided a logic and a mathematics compatible with it. And a significant source of understanding regarding the foundations of quantum mechanics then results.

\section*{Acknowledgments}
Our thanks go to Olival Freire Jr. for his help and support during the writing of this chapter.


\begin{thebibliography}{99}

\bibitem{are17} Arenhart, J. R. B. (2017). The Received View on Quantum Non-Individuality: Formal and Metaphysical Analysis. \textit{Synthese} \textbf{194}, pp. 1323-1347.

\bibitem{are19} Arenhart, J. R. B. (2019). Bridging the Gap between Science and Metaphysics, with a Little Help from Quantum Mechanics. In: J. D. Dantas, E. Erickson, and S. Molick (eds.) \textit{Proceedings of the 3rd Filomena Workshop}, pp.\,9-33. Natal: PPGFil UFRN.

\bibitem{are19a} Arenhart, J. R. B., Bueno, O. and Krause, D. (2019). Making Sense of Non-Individuals in Quantum Mechanics. In: O. Lombardi, S. Fortin, C. López, and F. Holik (eds.) \textit{Quantum Worlds: Perspectives on the Ontology of Quantum Mechanics}, pp. 185-204. Cambridge: Cambridge University Press.

\bibitem{bue14} Bueno, O. (2014). Why Identity is Fundamental. \textit{American Philosophical Quarterly}, \textbf{51}(4), pp. 325-332.

\bibitem{cha19} Chakravartty, A. (2019). Physics, Metaphysics, Dispositions, and Symmetries---\textit{à la} French. \textit{Studies in History and Philosophy of Science} \textbf{74}, pp. 10-15.

\bibitem{dal93} Chiara M.L.D., Di Francia G.T. (1993). Individuals, Kinds and Names in Physics. In: Corsi G., Chiara M.L.D., and Ghirardi G.C. (eds.) \textit{Bridging the Gap: Philosophy, Mathematics, and Physics}. Boston Studies in the Philosophy of Science, vol 140. Springer, Dordrecht.

\bibitem{cos07} da Costa, N.C.A., and Rodrigues, A.A.M. (2007). Definability and Invariance. \textit{Studia Logica} 86, 1–30.

\bibitem{bhk19} de Barros, A., Holik, F., and Krause, D. (2019). Indistinguishability and the Origins of Contextuality in Physics. \textit{Philosophical Transactions of the Royal Society A} \textbf{377}(2157) \url{https://doi.org/10.1098/rsta.2019.0150}.

\bibitem{die11} Dieks, D., and Lubberdink, A. (2011). How Classical Particles Emerge from the Quantum World. \textit{Foundations of Physics} 41 (6), 1051-1064.

\bibitem{dom10} Domenech, G., Holik, F., Kniznik, and Krause, D. (2010). No Labeling Quantum Mechanics of Indiscernible Particles. \textit{International Journal of Theoretical Physics} 49, 3085–3091.

\bibitem{dom08} Domenech, G., Holik, F., and Krause, D. (2008). Q-Spaces and the Foundations of Quantum Mechanics. \textit{Foundations of Physics} 38, 969–994.

\bibitem{esf19} Esfeld, M. (2019). Individuality and the Account of Nonlocality: the Case for the Particle Ontology in Quantum Physics. In: O. Lombardi, S. Fortin, C. López, and F. Holik (eds.) \textit{Quantum Worlds: Perspectives on the Ontology of Quantum Mechanics}, pp. 222-244. Cambridge: Cambridge University Press.

\bibitem{fre14} French, S. (2014). \textit{The Structure of the World: Metaphysics and Representation}. Oxford: Oxford University Press.

\bibitem{fre19} French, S. (2019). Identity and Individuality in Quantum Theory. In: E. N. Zalta (ed.) \textit{The Stanford Encyclopedia of Philosophy} (Winter 2019 Edition). URL = \url{https://plato.stanford.edu/archives/win2019/entries/qt-idind/}

\bibitem{fre06} French, S., and Krause, D. (2006). \textit{Identity in Physics: A Historical, Philosophical, and Formal Analysis}. Oxford: Oxford University Press.

\bibitem{jam74} Jammer, M. (1974). \textit{The Philosophy of Quantum Mechanics: The Interpretations of Quantum Mechanics in Historical Perspective}. New York: Wiley.

\bibitem{jec03} Jech, T. (2003). \textit{Set theory. The Third Millennium Edition, Revised and Expanded}. Berlin: Springer.

\bibitem{kra11} Krause, D. (2011). Is Priscilla, the Trapped Positron, An Individual? Quantum Physics, the Use of Names, and Individuation. \textit{Arbor} 187 (747): 61-66.

\bibitem{kra17} Krause, D., and Arenhart, J. R. B. (2017). \textit{The Logical Foundations of Scientific Theories: Languages, Structures and Models}. New York: Routledge.

\bibitem{kra18} Krause, D., and Arenhart, J. R. B. (2018). Quantum Non-Individuality: Background Concepts and Possibilities. In: Wuppuluri S. and Doria F. (eds.) \textit{The Map and the Territory}, pp. 281-305. Cham: Springer.

\bibitem{lad19} Ladyman, J. (2019). What is the Quantum Face of Realism? In: O. Lombardi, S. Fortin, C. López, and F. Holik (eds.) \textit{Quantum Worlds: Perspectives on the Ontology of Quantum Mechanics}, pp. 121-132. Cambridge: Cambridge University Press.

\bibitem{low03} Lowe, E. J. (2003). Individuation. In: Loux, M. J., and Zimmerman, D. W. (eds.) \textit{The Oxford Handbook of Metaphysics}. Oxford: Oxford University Press, pp. 75-95.

\bibitem{mau95} Maudlin, T. (1995). Three Measurement Problems. \textit{Topoi} \textbf{14}, pp. 7-15.

\bibitem{mit11} Mittelstaedt, P. (2011). The Problem of Interpretation of Modern Physics. \textit{Foundations of Physics} \textbf{41}(11), pp.1667-76.

\bibitem{mul08} Muller, F.A., and Saunders, S. (2008). Discerning Fermions. \textit{British Journal for the Philosophy of Science}, \textbf{59}, pp. 499-548.

\bibitem{pen07} Penrose, R. (2007). \textit{The Road to Reality: A Complete Guide to the Laws of the Universe}. New York: Vintage Books.

\bibitem{rue15} Ruetsche, L. (2015). The Shaky Game +25, Or: On Locavoracity. \textit{Synthese} \textbf{192}, pp. 33425-3442.

\bibitem{sch96} Schrödinger, E. (1996). \textit{Nature and the Greeks} and {Science and Humanism}. Cambridge: Cambridge University Press.

\bibitem{sch98} Schrödinger, E. (1998). What is an Elementary Particle? In: E. Castellani (ed.) \textit{Interpreting Bodies: Classical and Quantum Objects in Modern Physics}, pp. 197-210. Princeton: Princeton University Press.

\bibitem{shu17} Shumener, E. (2017). The Metaphysics of Identity: Is Identity Fundamental? \textit{Philosophy Compass}, doi:10.1111/phc3.12397.

\bibitem{skl10} Sklar, L. (2010). I'd Love to Ne a Naturalist---If Only I Knew What Naturalism Was. \textit{Philosophy of Science} \textbf{77}(5), pp. 1121-1137.

\bibitem{st02} Styer, D. F. et al. (2002). Nine Formulations of Quantum Mechanics. \textit{American Journal of Physics, Volume} 70(3), pp. 288-297.

\bibitem{sup60} Suppes, P. (1972). \textit{Axiomatic Set Theory}. (First edition, 1960.) New York: Dover.

\bibitem{tor81} Toraldo di Francia, G. (1981). \textit{The Investigation of the Physical World}. Cambridge: Cambridge University Press.

\end{thebibliography}
\end{document}